\documentclass[aps, prl, twocolumn, superscriptaddress]{revtex4}
\usepackage{graphicx}
\usepackage{dcolumn}
\usepackage{bm}
\usepackage{natbib}

\newcommand*{\cm}[1]{#1~cm$^{-1}$}

\begin{document}

\title{High-Frequency Electromagnon in GdMnO$_3$}

\author{A. M. Shuvaev}
\affiliation{Experimentelle Physik IV, Universit\"{a}t W\"{u}rzburg,
97074 W\"{u}rzburg, Germany} %
\author{F. Mayr}
\author{A. Loidl}
\affiliation{Experimentalphysik V, EKM, Universit\"{a}t Augsburg, 86135 Augsburg, Germany} %
\author{A. A. Mukhin}
\affiliation{General Physics Institute, Russian Acad. Sci., 119991
Moscow, Russia}
\author{A. Pimenov}
\affiliation{Experimentelle Physik IV, Universit\"{a}t W\"{u}rzburg,
97074 W\"{u}rzburg, Germany} %

\date{\today}

\begin{abstract}
We present the results of transmittance experiments on GdMnO$_3$
multiferroic manganite in the far infrared frequency range. The
spectra allow to obtain the positions and the intensity of the high
frequency electromagnon ($\sim$~\cm{75}) in this compound. We
present the comparative analysis of the high and low frequency
electromagnons across the phase diagram of GdMnO$_3$. The traces of
the electromagnon excitation can be detected even at room
temperature, i.e. deeply in the paramagnetic state.
\end{abstract}


\maketitle

Multiferroics are materials which simultaneously reveal magnetic and
electric order \cite{fiebig_jpd_2005, eerenstein_nature_2006}.
Enormous recent theoretical and experimental interest to these
systems is due to new interesting physics and possibilities for
applications especially in memory devices. Multiferroics are often
characterized by strong coupling between magnetism and electricity.
This can lead to such effects like switching of the electric
polarization in external magnetic fields or modulation of the
magnetization by electric field.

Dynamic properties of the multiferroics are quite rich as well. In
addition to classical modes of the ordered magnetic structure,
multiferroics show the existence of new excitations of
magnetoelectric nature \cite{pimenov_jpcm_2008, sushkov_jpcm_2008,
kida_josab_2009}. These excitations have been called electromagnons
and represent magnetic modes which can be excited by electric
component of the radiation. A very instructive example is
represented by the dynamics of TbMnO$_3$, because optical
experiments in this compound \cite{pimenov_nphys_2006,
takahashi_prl_2008, aguilar_prl_2009} can be combined with the
results by inelastic neutron scattering \cite{kajimoto_jpsj_2005,
senff_jpcm_2008}. Closely similar systems, like GdMnO$_3$ or
(Eu:Y)MnO$_3$ reveal in many aspects the same dynamic properties.
The electromagnon dynamics in these composition seem to consist of
two modes \cite{pimenov_jpcm_2008}. A strong high frequency
electromagnon is observed in the frequency range \cm{60-90}. The
frequency of this mode corresponds to a zone edge magnon as
experimentally proved in inelastic neutron scattering experiments
\cite{senff_jpcm_2008}. The spectral weight and excitation
conditions for this mode can well be explained
\cite{aguilar_prl_2009, miyahara_condmat_2008} by the Heisenberg
exchange mechanism combined with a modulation of a magnetic
structure. In addition, another electromagnons can be observed in
the frequency range of about \cm{23}. The origin of this mode, which
is often split into two modes, is not fully understood. The
characteristic positions of the low frequency electromagnon seems to
correspond to a magnon at the center of the magnetic Brillouin zone
\cite{senff_jpcm_2008, pimenov_prl_2009, shuvaev_prl_2010}. In these
scenario, the zone center magnons are electrically active modes of
the cycloidal magnetic structure and receive electric dipole
activity due to Dzyaloshinskii-Moriya mechanism
\cite{pimenov_jpcm_2008, katsura_prl_2007}. As this mechanism cannot
explain the intensity of the modes, the model including the magnetic
anisotropy has been recently suggested \cite{stenberg_prb_2009,
mochizuki_prl_2010}. A comparison of the model predictions with the
experimental parameters may help to resolve the problems with the
origin of electromagnons. Therefore, investigations of dynamical
properties of different multiferroics remain an actual task.

In this work we have carried out infrared transmittance experiments
on multiferroic GdMnO$_3$. The combination of these data with the
results by terahertz experiments \cite{pimenov_nphys_2006} allowed
to carry out a comparative analysis of low an high frequency
electromagnons in this material.

Single crystals of GdMnO$_3$ have been prepared using the
floating-zone method with radiation heating
\cite{balbashov_jcg_1996}. The samples have been characterized using
X-ray, magnetic and dielectric measurement
\cite{hemberger_prb_2004}. The basic properties of our samples agree
well with the results obtained by other groups \cite{goto_prl_2004,
kimura_prb_2005}. Transmittance spectra in the infrared frequency
range have been obtained using a Bruker IFS-113 Fourier-transform
spectrometer. For this purpose the ab-oriented cut of the crystal
was polished down to the thickness of 220 $\mu$m. Reflectance
spectra in the infrared frequency range have been measured on a
thick sample from the same batch and were published elsewhere
\cite{pimenov_prb_2006}. We note at this point that in spite of
relatively large electromagnon intensity, these modes are hardly
seen in reflectance spectra. Although the reflectance derived
spectra \cite{pimenov_prb_2006} do show some characteristic feature
around \cm{75}, without transmittance experiments the high frequency
electromagnon could not be observed unambiguously. In present work,
in order to obtain the full picture of the electromagnons, the
results by infrared transmittance are combined with the terahertz
spectra as obtained previously using the BWO-type technique
\cite{pimenov_nphys_2006}.

\begin{figure}[]
\includegraphics[width=0.9\linewidth, clip]{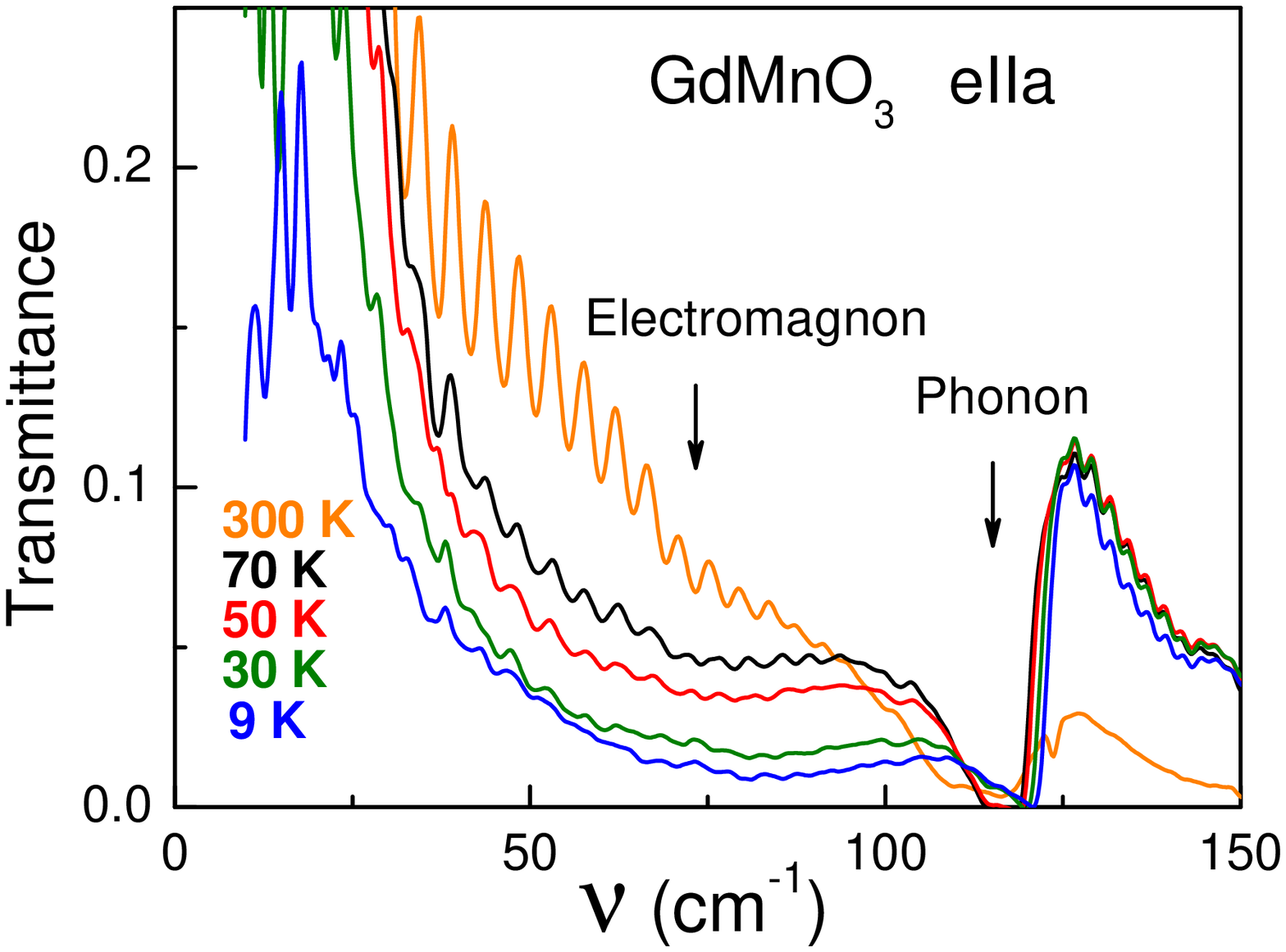}
%\vspace{0.2cm}
\caption{Transmittance spectra of a thin GdMnO$_3$ sample in the far
infrared frequency range and at different temperatures as indicated.
Arrows show the position of the high frequency electromagnon and of
the lowest infrared phonon. The low frequency electromagnon (around
\cm{23}) is not seen in this presentation.} \label{ftran}
\end{figure}

Figure~\ref{ftran} shows transmittance spectra of a thin GdMnO$_3$
sample in the far infrared frequency range. The strongest absorption
is observed close to \cm{120}, which corresponds to the lowest
phonon in GdMnO$_3$. Close to this phonon the transmittance is below
the sensitivity level of the spectrometer. Most importantly, a broad
minimum in transmittance can be seen around \cm{75} corresponding to
the high frequency electromagnon. Already at this point we may state
that  with increasing temperature the intensity of the electromagnon
gradually decreases which corresponds to the increase of the
transmission level. Even at room temperature, a weak local minimum
in transmittance close to \cm{75} can be seen indicating nonzero
electromagnon intensity at this temperature. The presented data are
sufficient to fill in the frequency gap between previous experiments
\cite{pimenov_nphys_2006, pimenov_prb_2006} in GdMnO$_3$, which gave
reliable spectra above \cm{100} and below \cm{40}.

\begin{figure}[]
\includegraphics[width=0.8\linewidth, clip]{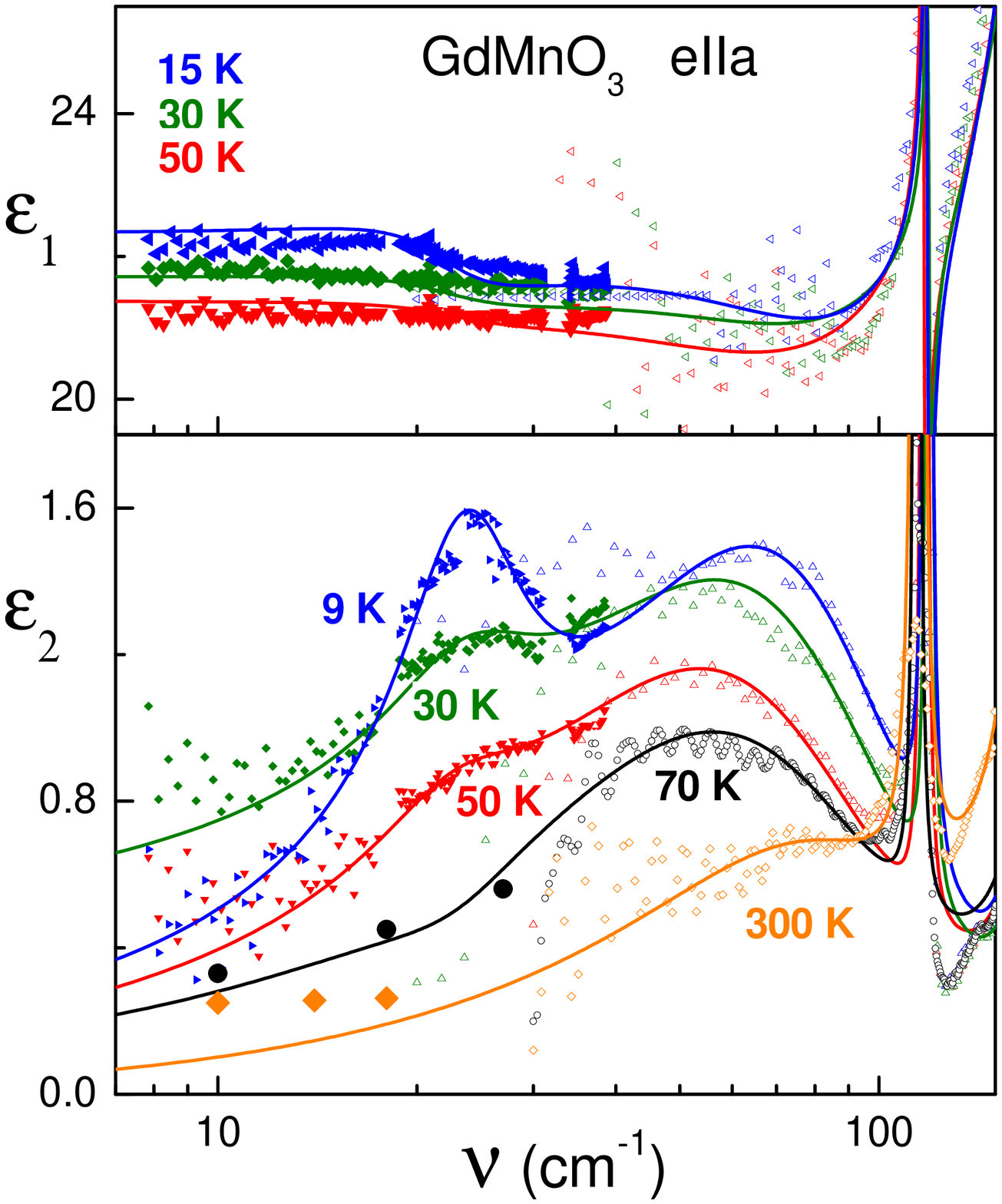}
%\vspace{0.2cm}
\caption{Complex dielectric permittivity of GdMnO$_3$ in the far
infrared frequency range. Open symbols - experimental data obtained
from transmittance and reflectance spectra, closed symbols - data
obtained from the complex transmission coefficient
\cite{pimenov_jpcm_2008, pimenov_nphys_2006}, solid lines - model
based on a sum of Lorentzians.} \label{feps}
\end{figure}

The measured transmittance spectra have been transformed to the
dielectric permittivity by inverting the Fresnel optical equations
for transmittance and reflectivity which neglect the interferences
within the sample. These interferences are seen as a Fabry-P\'{e}rot
type modulation of the transmittance spectra in Fig.~\ref{ftran} and
they are especially clear at room temperature and between \cm{20}
and \cm{60}. An attempt to take into account the interferences did
not improve the quality of the solution probably due to
imperfections of the sample surface. Figure~\ref{feps} represent the
far infrared spectra of the dielectric permittivity of GdMnO$_3$ in
the frequency range relevant for electromagnons. The results by the
infrared transmittance rapidly loose the accuracy below \cm{40}.
Therefore, the data by BWO spectroscopy \cite{pimenov_jpcm_2008,
pimenov_nphys_2006} are plotted as closed symbols in this frequency
range.

Two strong electromagnons can be well observed in the spectra of the
dielectric permittivity close to \cm{23} and \cm{75}. These modes
are seen most clearly in the imaginary part of the dielectric
permittivity (lower panel of Fig.~\ref{feps}). In the spectra of the
$\varepsilon_1$ (upper panel) only the low frequency electromagnon
can be detected. The reason of this effect is small dielectric
contribution ($\Delta \varepsilon \sim 0.5$) of the electromagnons
compared to the contributions of the phonons ($\Delta \varepsilon
\sim 20$). In order to obtain the parameters of both electromagnons,
the dielectric spectra in the far infrared frequency range were
fitted using the sum of several Lorentzians. Two low frequency
Lorentzians are responsible for the electromagnons and additional
higher frequency modes represent the contribution of the phonons. In
Fig.~\ref{feps} the effect of only two lowest phonons is important.

\begin{figure}[]
\includegraphics[width=0.99\linewidth, clip]{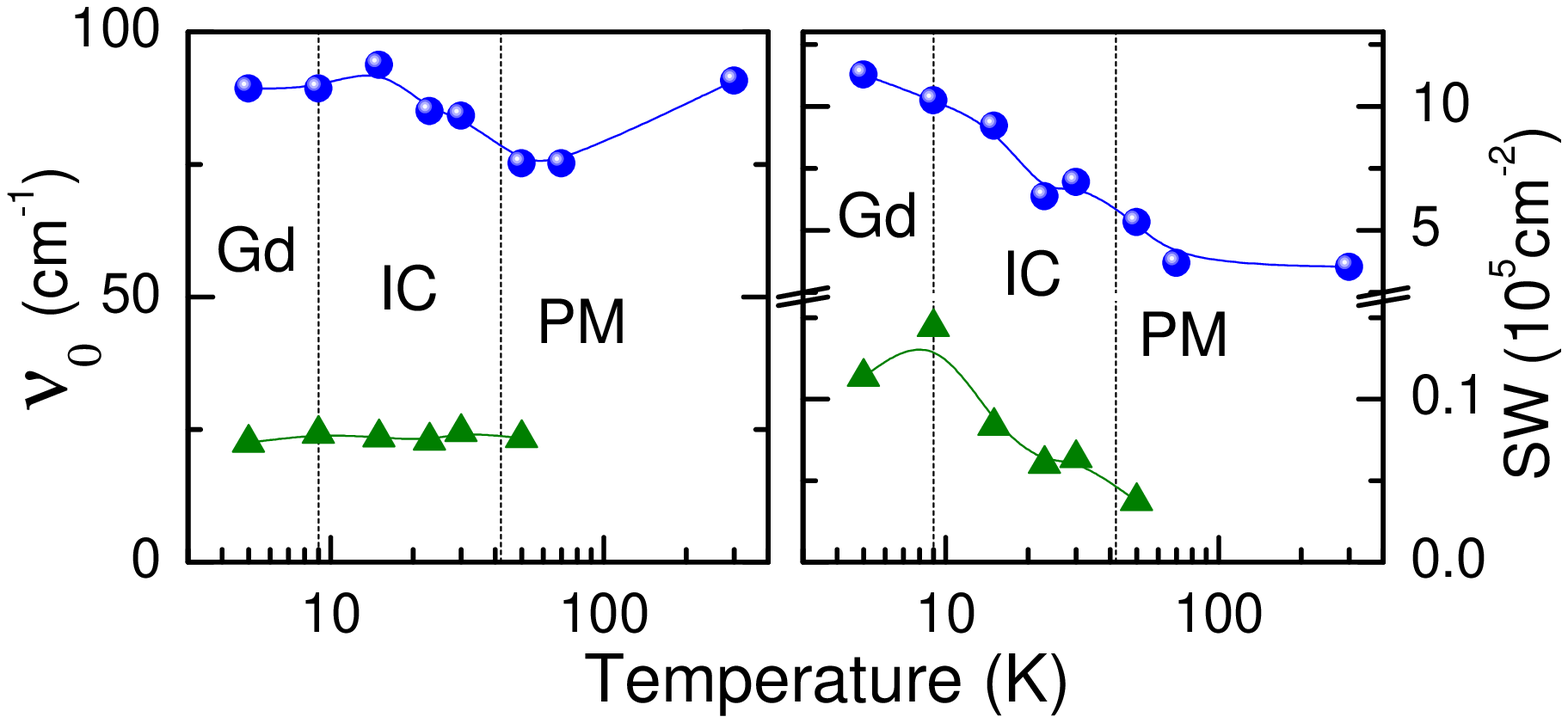}
%\vspace{0.2cm}
\caption{Parameters of the electromagnons in GdMnO$_3$. Left panel -
eigenfrequencies, right panel - spectral weight (note change in
scale in the vertical axis). Symbols - experimental data, lines are
to guide the eye. Thin dashed lines indicate phase boundaries
between different magnetic states \cite{kimura_prb_2005,
pimenov_prb_2006}: PM - paramagnetic, IC - incommensurate, Gd -
Gadolinium ordering. Note that in present experiments the canted
magnetic state is avoided because of zero external magnetic field
\cite{pimenov_prb_2006}. } \label{fpar}
\end{figure}

The dielectric parameters of both electromagnons are shown in
Fig.~\ref{fpar} as function of temperature. The resonance positions
as represented in the left panel reveal only weak temperature
dependence. As already mentioned, the high frequency electromagnon
can be observed even at room temperature, i.e. far above the
N\'{e}el phase transition. The eigenfrequency of this mode decreases
with the increasing temperature in the magnetically ordered state
and then slightly increases again towards room temperature. The
position of the low frequency electromagnon remains roughly
temperature independent. This mode cannot be detected in the spectra
above 50\,K. We note that the eigenfrequency especially of the high
frequency electromagnon in Fig.~\ref{fpar} is slightly higher than
the observed maximum in $\varepsilon_2$ (Fig.~\ref{feps}). This is
due to to the over-damped character of both electromagnons.
Especially the width of the high frequency electromagnon is quite
large $g= 125\pm5$\cm{} and, therefore, substantially shifts the
position of the maximum in $\varepsilon_2$.

Contrary to the rough temperature independence of the
eigenfrequencies, the spectral weight of the electromagnons in
GdMnO$_3$ is strongly temperature dependent. On heating the sample
into the paramagnetic state, the spectral weight of both
electromagnons decreases by about a factor of three. However, the
spectral weight of both electromagnons seem to remain finite even in
the paramagnetic state, which could probably be related to the
antiferromagnetic fluctuations.

Ar present, the nature of the high frequency electromagnon in
orthorhombic manganites seem to be settled \cite{aguilar_prl_2009,
miyahara_condmat_2008}. The frequency of this mode corresponds to
the zone edge magnon, the excitation conditions and the spectral
weight are well explained on the basis of the Heisenberg exchange
mechanism of the spin coupling. In order to provide an explanation
for the low frequency mode, a magnetic anisotropy
\cite{stenberg_prb_2009} and higher harmonics of the spin cycloid
\cite{mochizuki_prl_2010} within the Heisenberg exchange model have
been suggested. The ratio of the frequency positions of both
electromagnons $\nu_{0,1}/\nu_{0,2}=3.7 \pm 0.5$ approximately
agrees with the values in the (Gd:Tb)MnO$_3$ system
\cite{lee_prb_2009} and can be roughly accounted for by the
anisotropic model. On the contrary, the ratio of the spectral
weights for GdMnO$_3$ can be estimated as $0.011 \pm 0.002$. Such
small values agree with the tendency, observed in the (Gd:Tb)MnO$_3$
\cite{lee_prb_2009}, but differ by about an order of magnitude from
the theoretical estimates \cite{stenberg_prb_2009}. A possible
reason for this discrepancy is the absence of the true cycloidal
structure in GdMnO$_3$. In contrast to other orthorhombic manganite
multiferroics like TbMnO$_3$ or DyMnO$_3$, most probably no
cycloidal magnetic structure exists in the ordered state of
GdMnO$_3$. Another possibility can be suggested in analogy with the
discussion of the electromagnon mechanisms in TbMnO$_3$. In this
case the low frequency electromagnon simply corresponds to the zone
center magnon of the magnetic Brillouin zone \cite{senff_jpcm_2008,
pimenov_prl_2009, shuvaev_prl_2010}.

In conclusion, using far infrared  transmission spectroscopy, the
high frequency part of the electromagnon spectrum in GdMnO$_3$ has
been investigated. The spectral weight of the high frequency
electromagnon is roughly two orders of magnitude higher than that of
the low frequency mode. This ratio is the largest among other
orthorhombic manganite multiferroics and cannot be explained within
the existing models. The high frequency electromagnon can be seen in
the spectra even at room temperature, i.e. deep in the paramagnetic
state.

This work was supported by DFG (Pi 372).

\bibliography{literature}

\end{document}